\documentclass[%
reprint,
superscriptaddress,
amsmath,amssymb,
aps,
prb,
]{revtex4-2}
\usepackage{float}
\usepackage{color,graphicx}
\usepackage{ulem}
\graphicspath{ {images/} }
\usepackage{hyperref}
\begin{document}

\title{Single crystal growth of 1T-VSe$_2$ by molten salt flux method}

\author{Muhammed Anees K T}
\address{Department of Physics, Indian Institute of Technology Palakkad, KanjiKode, Palakkad, Kerala 678623, India}
\author{Souvik Kumar Rana}
\address{Department of Physics, Indian Institute of Technology Palakkad, KanjiKode, Palakkad, Kerala 678623, India}
\author{Abinash Das}
\address{Department of Physics, Indian Institute of Technology Palakkad, KanjiKode, Palakkad, Kerala 678623, India}
\author{Moumita Nandi}
\address{Department of Physics, Indian Institute of Technology Palakkad, KanjiKode, Palakkad, Kerala 678623, India}

\date{\today}

\begin{abstract}
VSe$_2$ is a highly promising van der Waals (vdW) material for applications in electronics, spintronics, and optoelectronics. Here, we report single crystal growth of 1T-VSe$_2$ by flux method using eutectic of NaCl/KCl molten salt. The typical size of as-grown VSe$_2$ single crystals is 5 $\times$ 4 $\times$ 0.1~mm$^3$. The elemental composition and homogeneity of the crystals were examined by energy dispersive x-ray spectroscopy, which is consistent with the stoichiometric ratio of VSe$_2$. The crystallographic [001] direction has been determined by x-ray diffraction. Raman measurement confirms that the 1T phase of VSe$_2$ has been formed. Temperature-dependent resistivity measurement exhibits a transition around 104 K due to the formation of the charge density wave phase.

\end{abstract}
\pacs{}
\keywords{}

\maketitle
\section{Introduction}
Two-dimensional (2D) van der Waals (vdW) materials provide ample opportunities to synthesize various potential candidates for applications in electronics, spintronics, and optoelectronics\cite{liu2016van,mak2016photonics}. Weak van der Waals forces make it easy to achieve low dimensionality by simply exfoliating bulk crystals. In addition to application purpose, van der Waals materials are equally important for fundamental studies. Among van der Waals materials, 2D layered transition metal dichalcogenides (TMDCs) exhibit a wide variety of novel complex phenomena such as superconductivity, charge density wave (CDW) phase, ferromagnetism, nontrivial topological bands, flat bands\cite{manzeli20172d,shi2015superconductivity,ritschel2015orbital,luxa2016origin, ma2019experimental,han2018van,wilson1975charge,wang2019higher,vitale2021flat}.

VSe$_2$, which belongs to the large class of the layered TMDC materials, provides a unique opportunity to realize all the above-mentioned novel phenomena in a single material. Additionally, VSe$_2$ has great potential in device applications. Band inversion and Dirac nodal arc have been observed by polarization-dependent angle-resolved photoemission spectroscopy (ARPES) experiments\cite{yilmaz2023dirac}. In addition, recent angle-resolved photoemission spectroscopy experiments reveal the presence of flat bands near the Fermi level, which makes VSe$_2$ proximal to the superconducting state\cite{yilmaz2022revealing}. Recently, Sahoo et al. have reported the emergence of superconductivity under high pressure\cite{sahoo2020pressure}. Apart from this, strong ferromagnetic ordering has been realized in the monolayer of VSe$_2$\cite{bonilla2018strong}. Most importantly, ferromagnetic ordering persists above room temperature, making VSe$_2$ a promising material for spintronics applications. In addition to this, the signature of the Kondo effect has been reported in VSe$_2$\cite{barua2017signatures}. Besides this, the charge density wave (CDW) phase has been observed in bulk\cite{strocov2012three}, as well as in monolayer\cite{duvjir2018emergence,wong2019evidence,chen2018unique}. Previous experiments have shown that VSe$_2$ adopts the 1T structure (octahedral) in the bulk form, whereas the 2H structure (trigonal prismatic) is more favored in the thin limit\cite{li2020structural}. Particularly, the 1T phase is metallic while the 2H phase exhibits semiconducting behavior\cite{li2020structural}. Moreover, 1T-VSe$_2$ is one of the rare correlated systems where the charge density wave ordering originates from three-dimensional (3D) nesting of the Fermi surface\cite{terashima2003charge}. In a nutshell, a diverse correlated physics properties can be studied in VSe$_2$. Although there are several extensive studies on VSe$_2$, some aspects are less explored such as the topological phase, superconductivity. Good quality single crystals are required to explore the physics in VSe$_2$. So far, single crystals of VSe$_2$ have been synthesized by the chemical vapor transport method using iodine as a transport agent. It is difficult to control the stoichiometric Selenium content in this method, and usually an unwanted vacancy is created at the chalcogen site.

In this work, we have reported single crystal growth of good quality 1T-VSe$_2$ by the molten salt flux method for the first time. We have characterized VSe$_2$ single crystals by energy dispersive x-ray spectroscopy and x-ray diffraction. The phase structure has been determined by Raman spectroscopy measurement. Our temperature-dependent resistivity measurement displays a clear signature of the charge density wave transition around 104 K. Therefore, the molten salt flux method provides an alternative way to obtain good quality single crystals of 1T-VSe$_2$.

\section{Experimental Methods}
Single crystals of VSe$_2$ were grown by the molten slat flux method. The raw materials of Vanadium powder (Thermo Scientific 99.5\%) and Selenium powder (Thermo Scientific 99.999\% ) with the stoichiometric amount of 1:2 molar \% were mixed with the salts NaCl/KCl flux. The molar ratio of the flux was NaCl:KCl = 1:1. The particular salt mixture was chosen for several reasons; first, it forms a low melting eutectic and remains liquid above 650 °C. Second, the resulting mixture is highly water-soluble; the crystals can be easily removed and cleaned by simply washing them with deionized water. The mixture of raw materials (0.6 g) and flux (3.6 g) was put in an alumina crucible and sealed in an evacuated quartz tube. The quartz tube was heated at 900 °C at a rate 30 °C/h and homogenized for 24 hrs, then cooled down to 650 °C at a rate of 2 °C/h followed by a furnace cooling to room temperature. The resulting crystals were washed with deionized water to remove the flux. Shiny plate-like single crystals formed. The typical size of the resulting bulk VSe$_2$ single crystals was 5 $\times$ 4 $\times$ 0.1~mm$^3$, shown is inset of Figure~\ref{Figure3}. The structural property was evaluated by x-ray diffraction (XRD). The as-grown single crystals were found to easily cleave along the crystallographic $ab$ plane, confirmed by the presence of only (00\textit{l}) diffraction peaks in the x-ray diffraction pattern. XRD measurement in a single crystal sample was performed using a Rigaku diffractometer with monochromatic Cu-K$_{\rm \alpha}$ radiation ($\lambda$ =0.15418 nm). The compositional ratio of the crystal was determined using energy dispersive x-ray spectroscopy (EDS). The EDS data were acquired using a scanning electron microscope (SEM) equipped with an EDS detector. Electrical transport measurement was performed using a Quazar Tech’s XPLORE 1.2 Physical Quantities Measurement System (PQMS). Raman measurement was performed in a HORIBA LabRAM HR Evolution Raman spectrophotometer set-up. A 532 nm laser beam was used to acquire the Raman data. Raman measurement was performed on the (001) plane of VSe$_2$ single crystal over the spectral range of 10-500 cm$^{-1}$.

\section{Results and Discussions}
\begin{figure}[t]
	\includegraphics[width=0.45\textwidth]{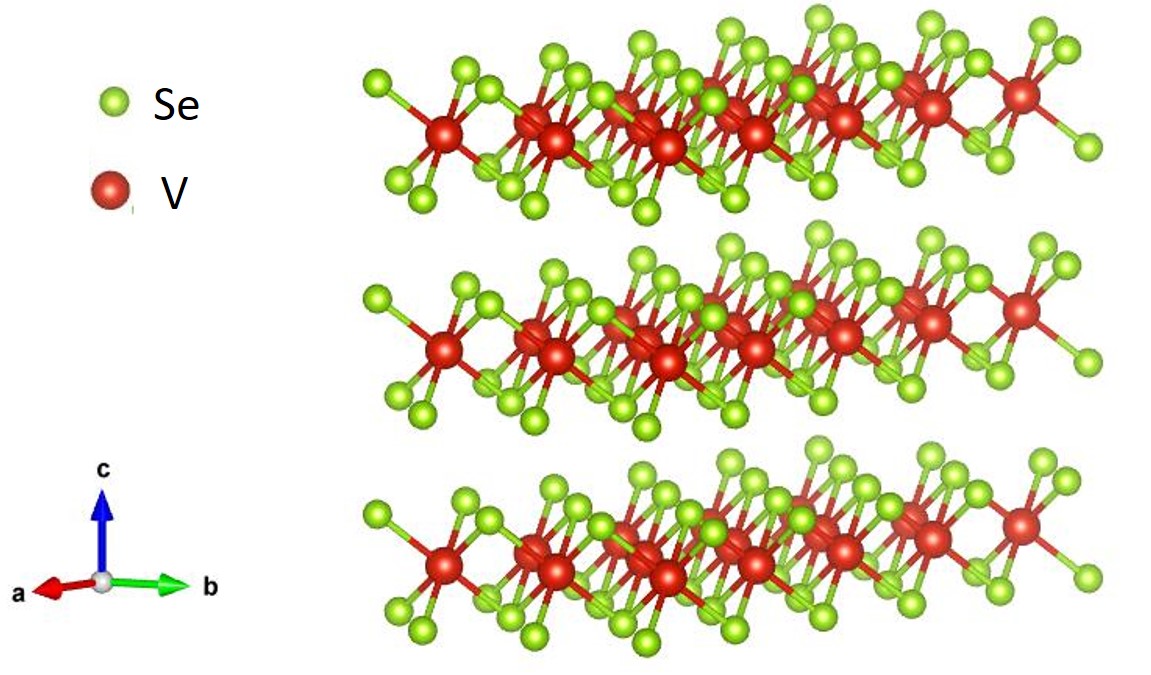}
	\caption{Crystal structure of VSe$_2$.}
	\label{Figure1}
\end{figure}

Figure~\ref{Figure1} presents the trigonal crystal structure of VSe$_2$, which crystallizes in the space group \textit{P}$\bar{3}$\textit{m}1. VSe$_2$ is a quasi-two-dimensional system with van der Waals interaction between layers. Here, the V atoms are covalently bonded to the octahedra of the Se atoms to form a single layer of VSe$_2$.

\subsection{Energy dispersive x-ray spectroscopy}
\begin{figure}[t]
	\includegraphics[width=0.5\textwidth]{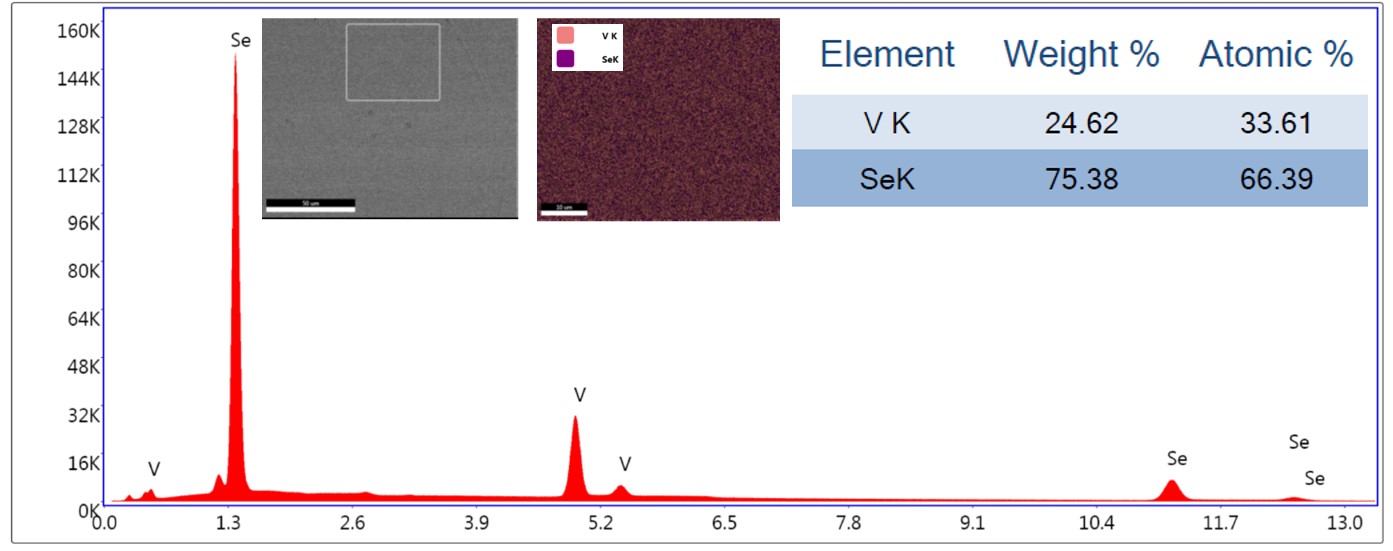}
	\caption{The EDS spectrum taken on the surface area of VSe$_2$ single crystal. The inset shows the SEM image taken on the flat surface of VSe$_2$ single crystal and elements V and Se distribution over the selected surface area.}
	\label{Figure2}
\end{figure}

Figure~\ref{Figure2} shows the EDS spectrum of the area scan of the VSe$_2$ single crystal piece. SEM images, shown in the inset of Figure~\ref{Figure2}, demonstrate that the composition of the crystal is quite homogeneous. The chemical composition is estimated to be V:Se = 1:1.98.\\ 
\subsection{X-ray diffraction}
\begin{figure}[b]
	\includegraphics[width=0.6\textwidth]{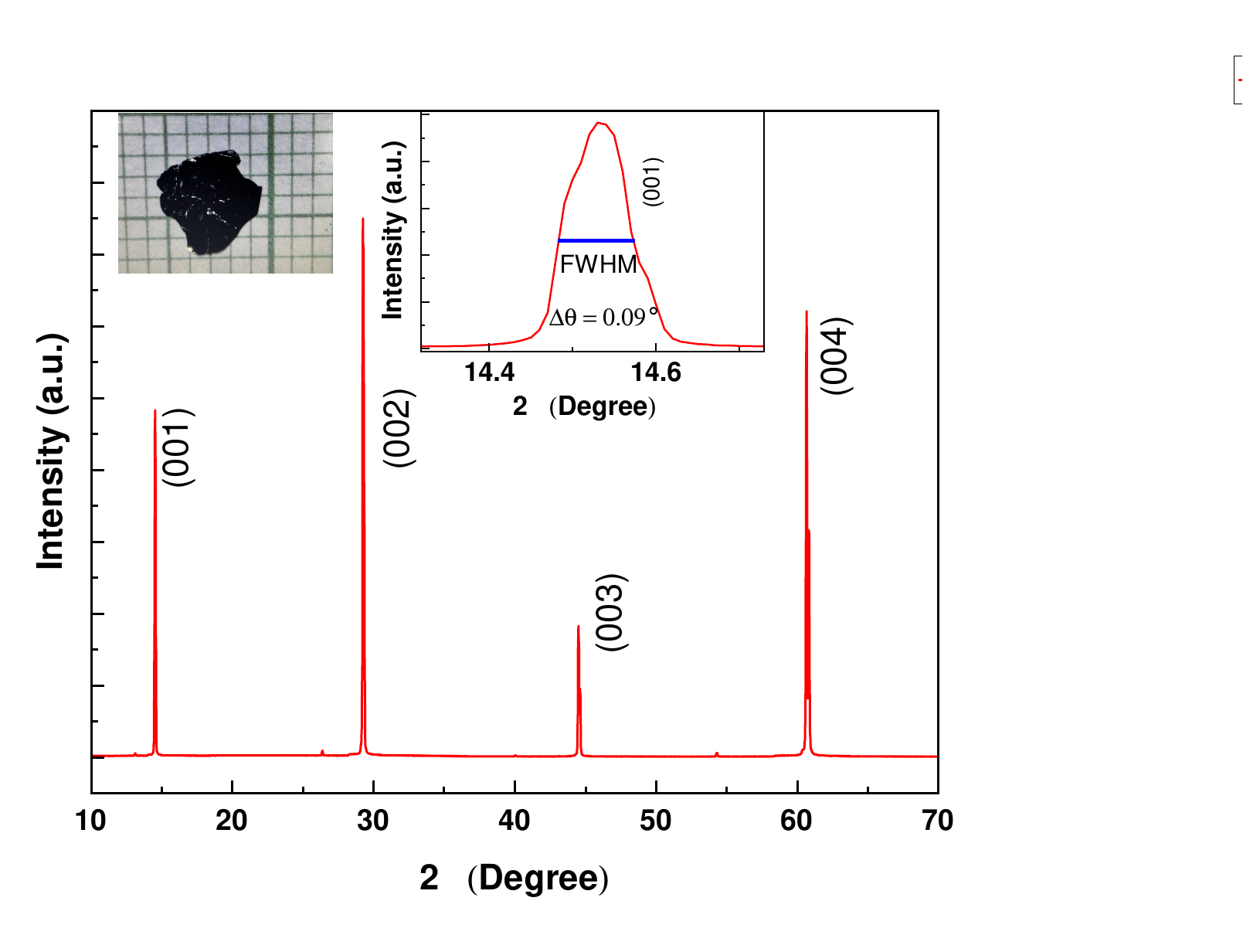}
	\caption{XRD pattern of the single crystal with (00\textit{l}) reflections; the inset of the right panel enlarges the (001) reflection.}
	\label{Figure3}
\end{figure}
Figure~\ref{Figure3} shows the XRD pattern of VSe$_2$
single crystal. The single crystalline nature
is confirmed by observation of the presence of only (00\textit{l}) diffraction peaks. The sharp peaks in the XRD $\theta$-2$\theta$ scan are indexed with (00\textit{l}) planes, indicating that the crystallographic $c$ axis is perpendicular to the surface of the plate, as shown in Figure~\ref{Figure3}. The inset of Figure~\ref{Figure3} shows that the full width at half-maximum (FWHM) of the (001) peak is only 0.09$^{\circ}$, indicating high crystalline quality. The value of the lattice parameter $c$ was extracted from these XRD data using Bragg's law. The extracted lattice parameter ($c$ = 6.09 \AA) is consistent with previous reports\cite{coleman1988scanning, bayard1976anomalous, wilson1969transition}.\\

\subsection{Raman spectroscopy}
\begin{figure}[t]
	\includegraphics[width=0.6\textwidth]{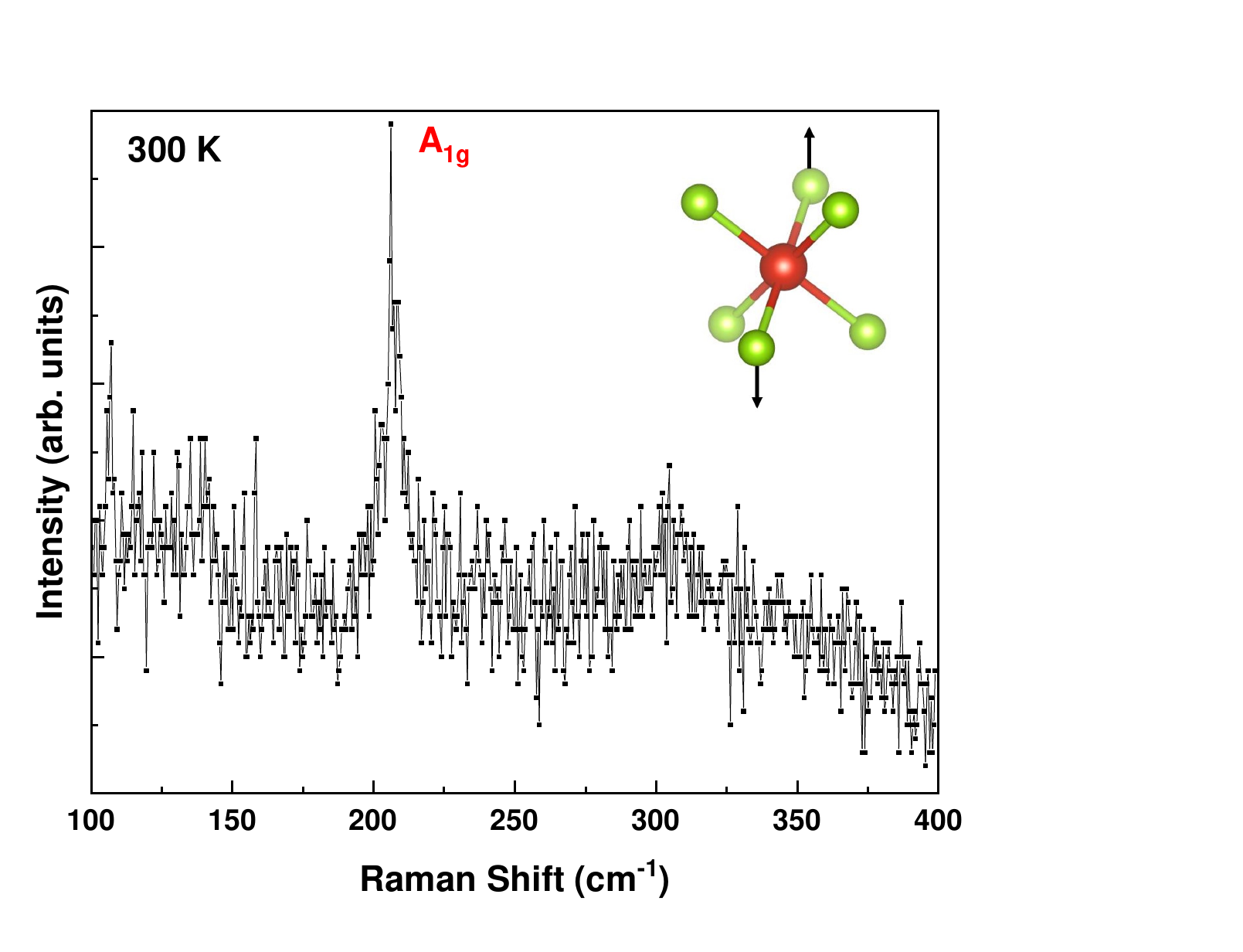}
	\caption{Raman spectra of 1T-VSe$_2$ crystal at 300 K; Inset shows schematic of crystal lattice vibration of A$_{1g}$ mode observed in Raman spectra.}
	\label{Figure4}
\end{figure}
To identify the phase structure, Raman spectroscopy was performed to examine Raman phonon modes at different positions. VSe$_2$ belongs to the space group $D_{3d3}$ and contains three atoms per unit cell\cite{goli2012charge,takaoka1980lattice}. There are nine zero-center vibrational modes for VSe$_2$, which can be presented with the following irreducible representation: 
\begin{equation}
\Gamma = A_{1g} + E_g(2) + 2A_{2u} + 2E_u(2)
\label{1}
\end{equation}
Here only two modes, $A_{1g}$ and the doubly degenerate $E_g$(2) modes, are Raman active while the other four are IR active modes\cite{ribeiro2014group}. In $A_{1g}$ mode two Se atoms per unit cell move relative to one another along the $z$-axis, while in $E_g$ mode the Se atoms move opposite to one another along the $x$ or $y$-directions. The Raman spectra presented in Figure~\ref{Figure4} display only one predominant peak at 206.1 cm$^{-1}$ in ambient conditions. The predominant peak at 206.1 cm$^{-1}$ corresponds to the $A_{1g}$ Raman active mode of 1T-VSe$_2$ which is consistent with earlier reports\cite{pandey2020electron,feroze2020depth,zhang2017van,yu2019chemically}. Recent DFT calculations for peak positions
of the Raman modes of 1T-VSe$_2$ also support our result\cite{wines2024combined}.

\subsection{Electrical transport properties}
\begin{figure}[t]
	\includegraphics[width=0.6\textwidth]{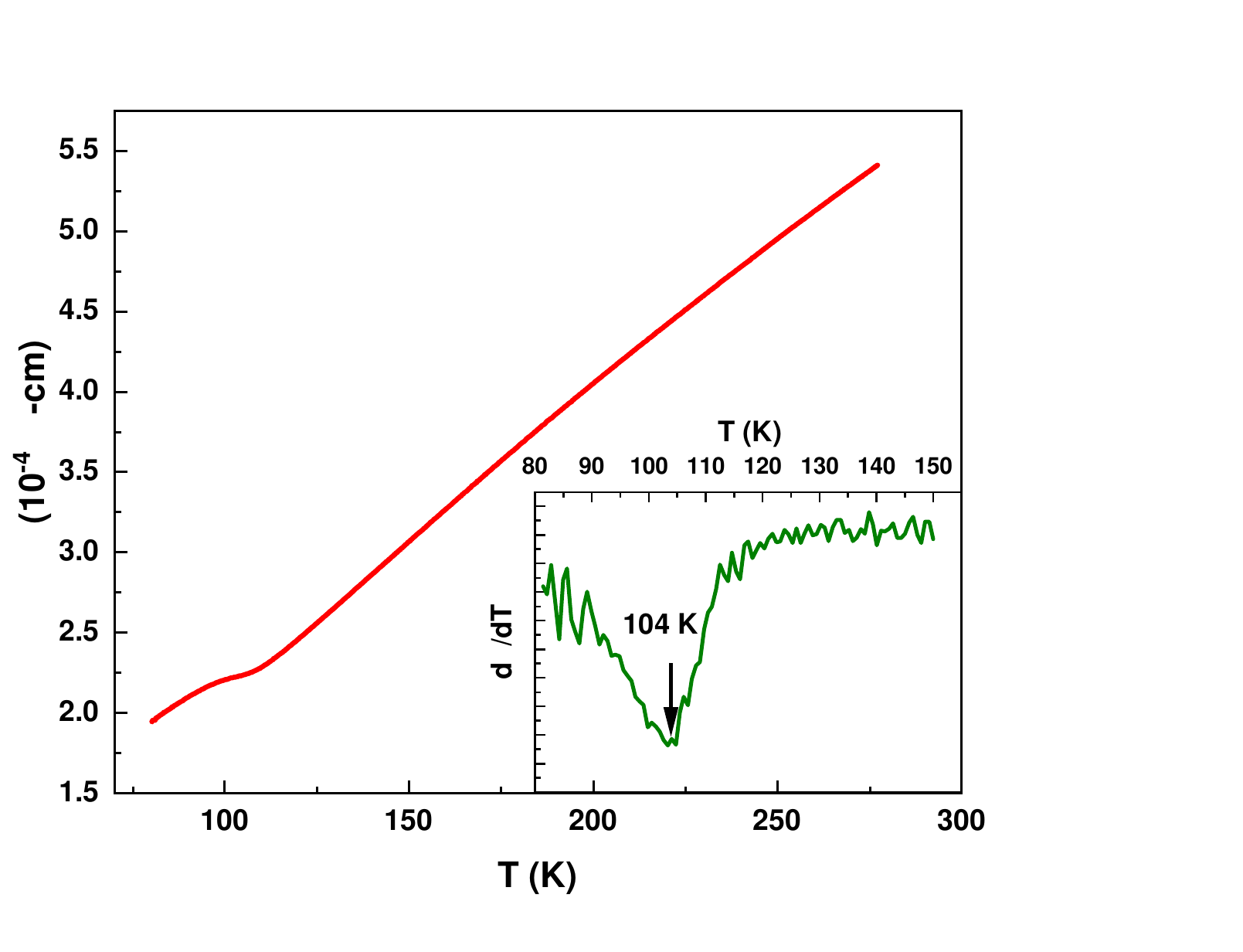}
	\caption{Temperature dependent resistivity ($\rho$ ) measurement at zero field; Inset shows derivative of resistivity vs $T$.}
	\label{Figure5}
\end{figure}

We have performed temperature-dependent resistivity measurement to investigate the electrical transport properties of 1T-VSe$_2$. Top layers of the sample were exfoliated using a scotch tape to avoid any possibility of oxidized surface. The sample was cut with a scalpel in a rectangular shape and mounted on electrically insulating substrates using GE varnish. Electrical contacts were made with the sample in a standard four-probe configuration with copper wires and conductive silver paint for resistivity measurement. Figure~\ref{Figure5} shows the temperature dependence of the longitudinal resistivity ($\rho$). Here, $\rho$ decreases with the decrease in temperature that is expected for the metallic nature of 1T-VSe$_2$. The plot $\rho$ vs. $T$ exhibits a kink at low temperature due to charge density wave transition. To evaluate the exact position of the transition, we have plotted the derivative of $\rho$, shown in the inset of Figure~\ref{Figure5}. The derivative of $\rho$ exhibits minima at 104 K, which is the charge density wave transition temperature of the 1T-VSe$_2$ single crystal. In support of our observation, K. Tsutsumi reported the periodic lattice distortion wave vector in the reciprocal lattice as (0.250 ± 0.003)a$^\ast$ +(0.314 ± 0.003)c$^\ast$ in the incommensurate CDW state below 110 K from temperature-dependent XRD measurements.\cite{tsutsumi1982x}.

\section{Conclusions}
For the first time, high-quality 1T-VSe$_2$ single crystals are successfully grown by the molten salt flux method. We have characterized the 1T-VSe$_2$ single crystals by XRD, EDS and Raman sectroscopy. The crystal structure and quality are confirmed by XRD. The rocking curve of the XRD pattern indicates that the single crystal is of high quality. Also, crystallographic $c$-direction has been determined by XRD. The Raman measurement confirms the formation of the 1T phase of VSe$_2$. The resistivity measurement exhibits the charge density wave transition around 104 K. Therefore, the molten salt flux method provides a new feasible scheme for growing high-quality single crystals of 1T-VSe$_2$.  

\section*{ACKNOWLEDGMENTS}
We acknowledge the Department of Science and Technology (DST) India (Innovation in Science Pursuit for Inspired Research-INSPIRE Faculty Grant). Also, we acknowledge the Central Instrumentation Facility, IIT Palakkad. We thank Ms. Reshna Elsa Philip for her help in performing energy dispersive x-ray spectroscopy measurements. We also thank Ms. Arathy Dileep for her help.

\bibliography{Ref}

\end{document}